# Interplay between the Kondo effect and randomness: Griffiths phase in $M_x$TiSe$_2$ ($M$ = Co, Ni, and Fe) single crystals

Minoru Sasaki[1], Akimasa Ohnishi[1], Takemasa Kikuchi[1], Mamoru Kitaura[1], Ki-Seok Kim[2], and Heon-Jung Kim[3*]

[1]Department of Physics, Faculty of Science, Yamagata University, Kojirakawa, Yamagata 990-8560 Japan

[2]Asia Pacific Centre for Theoretical Physics, POSTECH, Pohang, Gyeongbuk 790-784, Republic of Korea

[3]Department of Physics, College of Natural Science, Daegu University, Gyeongbuk 712-714, Republic of Korea

We investigate the interplay between the Kondo effect and randomness in $M_x$TiSe$_2$ ($M$ = Co, Ni, and Fe) single crystals. Although the typical low-T upturn of resistivity implies the Kondo effect around the single-ion Kondo temperature $T_K$, positive magnetoresistance linearly proportional to the magnetic field and the power-law scaling of magnetization suggest the forbidden coexistence between Kondo effect and time reversal symmetry breaking. This puzzling result is resolved by the Griffiths scenario—disorder-induced distribution of the Kondo temperature produces an effective Kondo temperature ($\overline{T}_K$) much lower than $T_K$, allowing unscreened local moments above $\overline{T}_K$ and resulting in non-Fermi liquid properties in $M_x$TiSe$_2$ below the percolation threshold (x<x$_c$).

Interplay between the Kondo effect and randomness is one of the central interests in modern condensed matter physics, such as for heavy fermion systems and dilute magnetic semiconductors, where non-Fermi liquid physics appears in both the magnetic properties and electrical transport beyond the Fermi liquid theory [1]. In particular, disorder-induced distribution of the Kondo temperature down to T = 0 K allows unscreened local moments, resulting in non-Fermi liquid physics—a typical example of the quantum Griffiths effect [2].

In this study we intercalate $TiSe_2$ with magnetic 3$d$ transition metals (Fe, Co, Ni) and investigate the possible quantum Griffiths phase from both magnetic and transport properties. Layered compounds Ti$X_2$ ($X$ = $S$, $Se$, $Te$ : chalcogen) are well known for their low-dimensional electronic structures that are easily intercalated with guest atoms [3,4], which produce two-dimensional semiconductors ($TiS_2$), semimetals ($TiSe_2$), and metals ($TiTe_2$), depending on the degree of electron transfer from the chalcogens to Ti 3$d$ orbitals and of the $p$-$d$ hybridization. In particular, $TiSe_2$ displays an unexpected charge density wave (CDW) beyond the conventional nesting picture, where the Fermi surface consists of six ellipsoidal pockets. It is fascinating that Cu intercalated $TiSe_2$ ($Cu_xTiSe_2$) exhibits superconductivity around 4.2 K with doping dependence similar to that of $T_c$ with high-$T_c$ cuprate superconductors [5].

When magnetic 3$d$ transition metals ($M$ = Co, Ni, Fe) are intercalated into $TiSe_2$ ($M_xTiSe_2$), they act as spin-flip scattering centers to produce the Kondo effect in the dilute limit. Increasing the concentration of intercalated magnetic impurities, their random distribution is expected to cause quantum Griffiths effects in the disordered Kondo system [1]. When the concentration is above a critical value ($x_c$), interpreted as the percolation threshold, Ruderman-Kittel-Kasuya-Yosida (RKKY) interactions between the intercalated local quantum spins become important in competing with the Kondo effect, providing a framework to describe heavy fermion physics [2]. Although this is the desired overall picture, the expected non-Fermi liquid physics is rare in the disordered Kondo regime (x < $x_c$). $M_xTiSe_2$ with magnetic $M$ provides an opportunity to examine the disordered Kondo effect systematically.

In this communication we investigate the interplay between the Kondo effect and randomness in $M_xTiSe_2$ ($M$ = Co, Ni, and Fe) single crystals. Although the typical low-T upturn of resistivity implies the Kondo effect around the single-ion Kondo temperature $T_K$, positive magnetoresistance (MR) linearly proportional to the magnetic field and power-law scaling of magnetization deviates from the local Fermi liquid picture, suggesting the forbidden coexistence of the Kondo effect and time reversal symmetry breaking. We resolve this puzzling result, based on the quantum Griffiths scenario that disorder-induced distribution of the Kondo temperature causes an effective Kondo temperature ($\bar{T}_K$) much lower than $T_K$, allowing unscreened local moments above $\bar{T}_K$ and resulting in non-Fermi liquid properties in $M_x$TiSe$_2$ below the percolation threshold (x< $x_c$ ≈ 0.07). On the other hand, $M_x$TiSe$_2$ evolves into the usual metallic phase above $x_C$.

In the previous work on Fe$_x$TiSe$_2$ [6], the peak in ρ(T) curves, characterizing the CDW, decreased appreciably when increasing the intercalated Fe concentration $x$ up to the critical concentration $x_c$ (0.065 < $x_c^{exp}$ < 0.075), leaving anomalously high resistivity at low temperature—above $x_c$, the transport properties change dramatically and a metallic state is recovered. The critical concentration $x_c$ is reported to be a point where a percolation path of Fe clusters forms. According to an ARPES study [7], when the concentration of $x$ is high, Fe intercalation produces a nondispersive impurity-induced band near the Fermi level. This strongly suggests that the intercalated Fe acts as a dopant and also as an impurity. Therefore, well below $x_C$, the high resistivity at low temperature is believed to result from single impurity scattering. In fact, qualitatively similar behavior is observed in $M_x$TiSe$_2$ ($M$=Ni and Co), but not in Cu$_x$TiSe$_2$, as will be discussed below.

The single crystals of the 3d transition metal $M$ intercalation compound $M_x$TiSe$_2$ [$M$ = Ni; $x$ ≤ 0.089 (> $x_c$), $M$ = Co; $x$ ≤ 0.13 (> $x_c$), and $M$ = Cu; $x$ ≤ 0.06 (< $x_c$)] were grown by a chemical vapor transport technique in the presence of iodine as a transport agent [6,8]. To avoid co-intercalation of guest $M$ atoms and constituent Ti atoms into the host TiSe$_2$, all compounds were grown at relatively low temperature (500°C). Above that temperature, Ti atoms are known to self-intercalate into the host [9]. Values of the intercalated guest concentration $x$ were determined by electron-probe microanalysis. X-ray powder diffraction measurements for $M_x$TiSe$_2$ indicated that these compounds have a 1$T$-CdI$_2$ structure. Resistivity measurements were performed in the temperature range from 4.2 to 300 K using a DC four-probe method. MR measurements were carried out at 4.2 K and at magnetic fields up to 4.0 T. Zero-field cooled (ZFC) magnetizations were investigated with a SQUID magnetometer at magnetic fields of 0.02 and 0.2 T. In MR and magnetization measurements, magnetic fields were applied perpendicular to the layer plane.

Figure 1(a) shows the temperature dependence of a resistivity component ρ$_A$ of Cu$_x$TiSe$_2$, whose phonon and residual contributions are subtracted. The detailed method for analysis and raw data are presented in the supplementary information [10]. The phonon and residual contributions are estimated as $\rho_{bg} = \alpha\rho_r$, where $\rho_r$ is the resistivity curve above $x_c$ and $\alpha$ is the scaling parameter with nearly unity value. In the samples at x>$x_c$, the main contribution for resistivity are the electron-phonon and electron-electron scatterings in high and low T regions, respectively. Although the extracted resistivity slightly depends on the parameter $\alpha$, the main features are retained. As can be seen in Fig. 1(a), this method isolates resistivity peaks for $Cu_xTiSe_2$. As the concentration of x increases, the height of the peak is reduced and the peak moves to lower temperatures. However, the shape of the peak, especially below the maximum value, remains unchanged. The scaled curves are shown in fig. S1 in the supplementary information [10]. Fig. 1(a) clearly shows that the CDW is suppressed with increasing x, which is consistent with previous reports. Quantitatively, this decrease is in good agreement with Ref. [5] but not Ref. [11], where the CDW is more rapidly suppressed. Similar CDW suppression was reported in $Fe_xTiSe_2$. Below the peak region, $\rho_A$ is quite small in $Cu_xTiSe_2$.

We applied the same analysis to $M_xTiSe_2$ (*M*=Fe, Ni, and Co). As an example, we present the results for $Co_xTiSe_2$ in Fig. 1(b)—the results are similar for the other compounds. This figure clearly shows the difference for $Cu_xTiSe_2$. The low-T upturn of resistivity exists in $M_xTiSe_2$ (*M*=Fe, Ni and Co). Since this low-temperature part is observed only in $M_xTiSe_2$ (*M*=Fe, Ni, and Co), it is not related to the CDW. At reasonably high x at x<$x_c$, the CDW peak is comparatively small and it is quite clear that the low-temperature part becomes dominant with high x. Since the peak shape is independent of x for $Cu_xTiSe_2$, we subtracted this peak numerically, assuming that the shape of CDW peaks is same in all $M_xTiSe_2$ compounds, especially below the maximum. We could get a low-T upturn of resistivity with a clear shape. The low-T upturns at different values of x share common features, as shown in Fig. 1(c). This low-temperature component increases with decreasing temperature, approximately following a logarithmic dependence in *T* and, saturated with $T^n$ dependence at low temperature as shown in Fig. 1(d). This overall trend is reminiscent of spin flip scattering and spin singlet formation in the dilute Kondo system. For the $M_xTiSe_2$ compound with nearly independent spins at x<$x_c$, it is quite reasonable to expect the Kondo effect to appear.

If the Kondo effect is the origin of the anomalous increase in resistivity at low temperature, its effect should be manifested in other measurable quantities, such as MR and magnetization. In the Kondo system, when spin flip scattering is dominant, negative MR should appear, because of the suppression of spin flip scattering by magnetic fields *B*. On the other hand, at lower temperatures, when spin singlet is formed, the usual positive *B*-quadratic MR is anticipated [12,13]. However, what we observe is quite different from these predictions. For $Cu_xTiSe_2$, a positive *B*-quadratic MR is observed, as shown in Fig. 2(a), implying that MR in this case results from the reduction of the mean free path by cyclotron motion. However, for $M_x$TiSe$_2$ ($M$= Ni and Co), MR is dominantly positive and linear with $B$ at x<x$_c$, whereas it is purely quadratic at x>x$_c$. In fact, MR is quadratic only in low fields and becomes linear in the region of x<x$_c$. This crossover occurs at around 1 T or less and is characterized by the crossover field $B_{cross}$. The quadratic MR results from impurity scattering and cyclotron motion of charge carriers.

According to recent theoretical investigations [14], positive linear MR can be observed in the low field region when the Kondo temperature $T_K$ distributes down to 0 K; thus, spins are unquenched even at the zero temperature limit. The MR changes to negative in high fields in which the Zeeman energy exceeds the energy scale set by the peak value of $T_K$. Note that our maximum field of 4 T corresponds to a temperature of 2.4 K, while the temperature where the resistance starts to increase is on the order of 100 K. This verifies that our MR measurements are performed in the low field region.

The natural consequence of the above scenario is the existence of unquenched spins, even at low temperature. In this case, magnetization should follow the power-law behavior of

$$M \sim \int_0^\infty P(T_K)(T - aT_K)^{-1} dT_K \sim (T - \theta)^{\alpha - 1},$$

where $P(T_K)$ is the distribution function of $T_K$, and $\alpha$ is the exponent, which depends on $P(T_K)$, and $\theta$ is a small quantity [15]. The $\alpha$ value becomes smaller as the disorder strength W/t increases, where W is the band width broadened by disorder and t is the hopping integral [15]. This power law behavior is quite different from what is expected in the usual dilute Kondo system. In the dilute Kondo system, magnetization is suppressed below $T_K$ because of the formation of a spin singlet of an impurity spin and conduction electrons. In $M_x$TiSe$_2$ ($M$=Fe, Ni, and Co), the former behavior is observed, as shown in Fig. 3, suggesting that spins remain unquenched up to a certain temperature—much lower than the Kondo temperature of the resistivity data. This implies a huge distribution of the Kondo temperature. On the other hand, Cu$_x$TiSe$_2$ follows a simple Curie behavior. The value of $\alpha$, which is obtained from fitting data in the region of 4 K < $T$ < 80 K, is around 2. According to the calculations of Miranda and Dobrosavljevic [15], this value corresponds approximately to a disorder strength W/t of 0.8. Our $\alpha$ values are not far from the critical value ($\alpha$=1), where disorder-driven non-Fermi liquid behavior is expected. The determination of a more precise $\alpha$ value and of the degree of the criticality needs more precise measurements at $T$ < 4.2 K.

To understand how electrical transport properties change across $x_c$, we present the scattering cross section $A$ of Fe$_x$TiSe$_2$ in $\rho(T) \sim AT^2$ at $x > x_c$, the crossover field $B_{cross}$ in MR, and $\Delta\rho/\rho$ at 4 T in Figs. 4(a), (b), and (c), respectively. As in Fe$_x$TiSe$_2$, above $x_c$, the resistivity follows a metallic behavior in $M_x$TiSe$_2$ ($M$= Ni, and Co). It is very interesting that this metallic behavior exists only when a percolation path is formed and this implies that a coherent effect become important in stabilizing this metallic state. Moreover, as seen in Fig. 4(a), A becomes larger, approaching $x_c$. This suggests that the electron-electron scattering becomes stronger near $x=x_c$.

This Fermi liquid state is abruptly destroyed below $x_c$. In fact, this change accompanies an evolution of MR, as shown in Fig. 4(b). The region of $B$-linear MR is quite robust at $x<x_c$ but such a region does not exist at $x>x_c$. In contrast, Cu$_x$TiSe$_2$ shows only a positive $B$-quadratic MR. This is quite reasonable because the $B$-linear term in MR originates from the Kondo disorder mechanism, while the quadratic term results from impurity scattering and the cyclotron motion of charge carriers. Therefore, the large region of linear MR in $M_x$TiSe$_2$ ($M$= Ni, and Co) implies that the Kondo effect with huge $T_K$ distribution is important at $x<x_c$. This effect disappears concomitant with the formation of a percolation path at $x=x_c$.

The positive $B$-linear MR with the power-law scaling of magnetization implies time reversal symmetry breaking, at least locally in time, associated with unscreened local moments in the disordered Kondo regime; it seems to recover over longer time scales to cause the typical local Fermi liquid picture, associated with the Kondo effect. This leads us to introduce an effective Kondo temperature $\overline{T}_K$ that is much lower than the single-ion Kondo temperature $T_K$, originating from the disorder-induced distribution $P(T_K)$ of the Kondo temperature. The Kondo effect in resistivity appears around $T_K$; however, there are some clusters with unscreened local moments due to randomness (called rare regions in the Griffiths scenario) giving rise to unexpected non-Fermi liquid behavior above the disorder-average Kondo temperature. This interpretation is completely consistent with the crossover magnetic field $B_{cross}$, identified with $B_{cross} \approx \overline{T}_K$, where the $B$-linear MR results when $B>B_{cross}$, while the typical quadratic B dependence appears below this energy scale, recovering the local Fermi liquid.

This situation changes drastically above $x_c$. When the percolation path is formed near $x_c$, the disorder-average Kondo temperature is not much different from the single-ion Kondo temperature. As a result, the regime of time reversal symmetry breaking narrows, and the positive $B$-linear MR disappears with the Kondo effect. RKKY correlations can compete with the Kondo effect above the percolation threshold, but our experiment tells that the RKKY energy scale is smaller than the Kondo temperature, and not relevant in low energy physics.

It is quite interesting that the magnitude of MR also shows discontinuous change near $x_c$, as shown in Fig. 4(c). However, we do not understand how it changes, as the number and size of impurity clusters increase at this moment. Since DC magnetization probes the time average of Kondo fluctuations, a power-law behavior in magnetization cannot be so pronounced.

Next, the strength of the Kondo disorder effect and different $\alpha$ values should be discussed. Experimentally, this effect is most pronounced in $Fe_xTiSe_2$ among the $M_xTiSe_2$ compounds (*M*=Fe, Ni, and Co), and yields the largest low-T upturn of resistivity. This compound even shows a variable-range power-law increase of resistivity in the low-T region, around x= 0.05. These observations suggest that $Fe_xTiSe_2$ near $x_c$ is very close to a non-Fermi liquid state. Two scenarios might describe the origin of this situation: First, as pointed out by Miranda and Dobrosavljevic [1], the distribution of Kondo impurities can be different in a different $M_xTiSe_2$; therefore, the strongest effect in $Fe_xTiSe_2$ is due to the critical distribution of $T_K$, whereas the others are relatively far from the critical distribution. Different distributions of Kondo impurities are possible only when the chemistry of the *M* cluster formation is quite different. The second possibility is the enhanced Kondo fluctuations due to RKKY interaction. This effect is believed to be strongest in $Fe_xTiSe_2$ because Fe has the largest spin and thus, the strongest correlation between Fe spins, making Kondo fluctuations more critical in $Fe_xTiSe_2$. Because the chemistry is not thought to be so different in the low concentration region, the former scenario is more plausible. In contrast, the distribution of Kondo impurities could be important for the series with fixed M in $M_xTiSe_2$, and this may be the origin of the observation that the low-T upturn is largest at x=0.019 in $Co_xTiSe_2$.

Finally, it is interesting to compare linear MR of the $Ag_{1+\delta}Se$ and $Ag_{1+\delta}Se$ [16] with the present case. Even though quantum magnetoresistance [17] and strong inhomogeneity [18] have been proposed as possible origins for the large linear MR, these mechanisms cannot be applied to the present case in respect that they do not consistently explain the low-T upturn in resistivity and the power-law behavior of magnetization.

The interplay between the Kondo effect and randomness seems to allow the forbidden coexistence of Kondo effect and time reversal symmetry breaking in $M_xTiSe_2$ (*M*=Fe, Ni, and Co) single crystals, as indicated by a positive magnetoresistance that is linearly proportional to the magnetic field. This unexpected result is naturally resolved, introducing an effective Kondo temperature $\bar{T}_K$ much lower than the single-ion Kondo temperature $T_K$ due to the disorder average. This quantum Griffiths scenario allows linear magnetoresistance when

$T_K > T > \bar{T}_K$, while the local Fermi liquid results at low temperatures $\bar{T}_K > T$, are completely consistent with the crossover magnetic field $B_{cross} \approx \bar{T}_K$, where the typical $B^2$ dependence arises when $B_{cross} > B$. More detailed investigations are needed near the percolation threshold, and are expected to show interplay between the quantum criticality and randomness.

**ACKNOWLEDGEMENTS**

This research was supported by National Nuclear R&D Program through the National Research Foundation of Korea (NRF) funded by the Ministry of Education, Science and Technology (20100018383).

* Corresponding author : hjkim76@daegu.ac.kr

**Figure captions**

Fig. 1 (Color online) Anomalous increase of resistivity $\rho_A$ of (a) $Cu_xTiSe_2$ and (b) and $Co_xTiSe_2$. The background is subtracted as explained in the text. The $\rho_A$ of $Cu_xTiSe_2$ consisting solely of a CDW peak and the low-$T$ part of $\rho_A$ is almost zero. In contrast, the low-$T$ part of $\rho_A$ is considerably high in $Co_xTiSe_2$. The red arrow indicates the CDW transition temperature. (c) The $\rho_A$ is decomposed into a CDW peak ($\rho_{AP}$, red curve) and a low-$T$ upturn ($\rho_{mL}$, green curve). (d) The temperature dependence of $\rho_{mL}$ of $Co_xTiSe_2$. $M_xTiSe_2$ ($M$ = Fe and Ni) all exhibit similar behavior.

Fig. 2 (Color online) The MR of (a) $Cu_xTiSe_2$, (b) $Co_xTiSe_2$, and (c) $Ni_xTiSe_2$ measured at T=4.2 K. While the MR of $Cu_xTiSe_2$ is quadratic with $B$, the MR of other compounds shows a dominantly linear dependence over a wide B range at $x<x_c$. In the low field region, this linear MR changes into quadratic—only quadratic MR appear at $x>x_c$.

Fig. 3 (Color online) Inverse magnetic susceptibility $\chi^{-1}$ = H/M of $M_xTiSe_2$ ($M$=Fe, Ni, Co, and Cu). While $Cu_xTiSe_2$ follows a Curie behavior, $M_xTiSe_2$ ($M$=Fe, Ni, and Co) shows a power-law dependence. The solid lines are theoretical fit with the formula $M \sim\sim (T-\theta)^{\alpha-1}$. The values of $\alpha$ are 1.92, 1.86, and 2.3~2.4 for $Fe_xTiSe_2$, $Ni_xTiSe_2$, and $Co_xTiSe_2$, respectively. Those of $\theta$ are -2, 0, and -13 K.

Fig. 4 (Color online) (a) The concentration of x dependence of A values in $Fe_xTiSe_2$; (b) the crossover field $B_{cross}$ of $M_xTiSe_2$ ($M$=Fe, Ni, Co, and Cu) from quadratic to linear dependence; and (c) $\Delta\rho/\rho$ at 4 T (c). These quantities change character at $x\sim x_c$ ($x_c \sim 0.07$) and this critical concentration is indicated by the dotted line. The grey part is the experimentally-determined critical region ($0.065 < x_c^{exp} < 0.075$), where a percolation path is known to be formed [6].

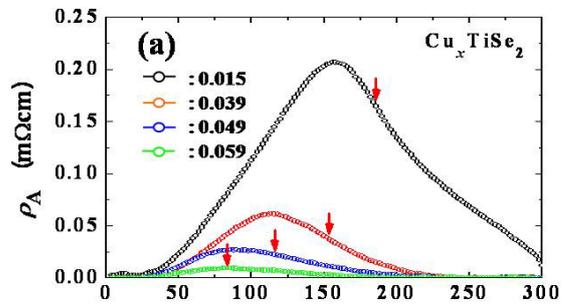
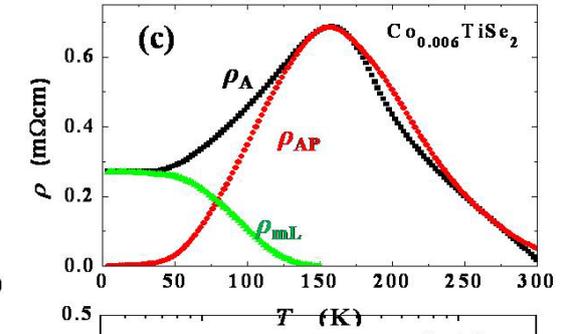
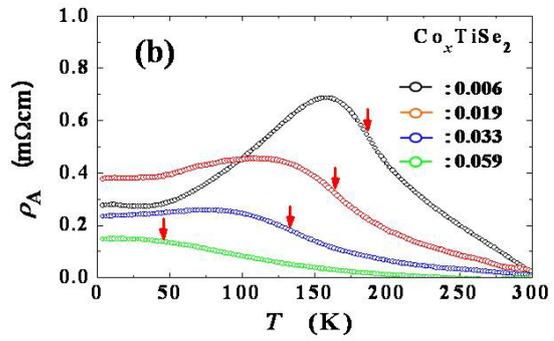
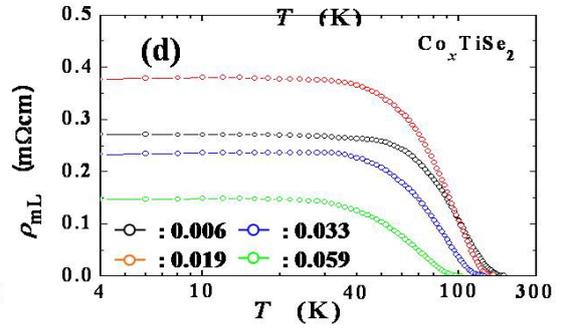

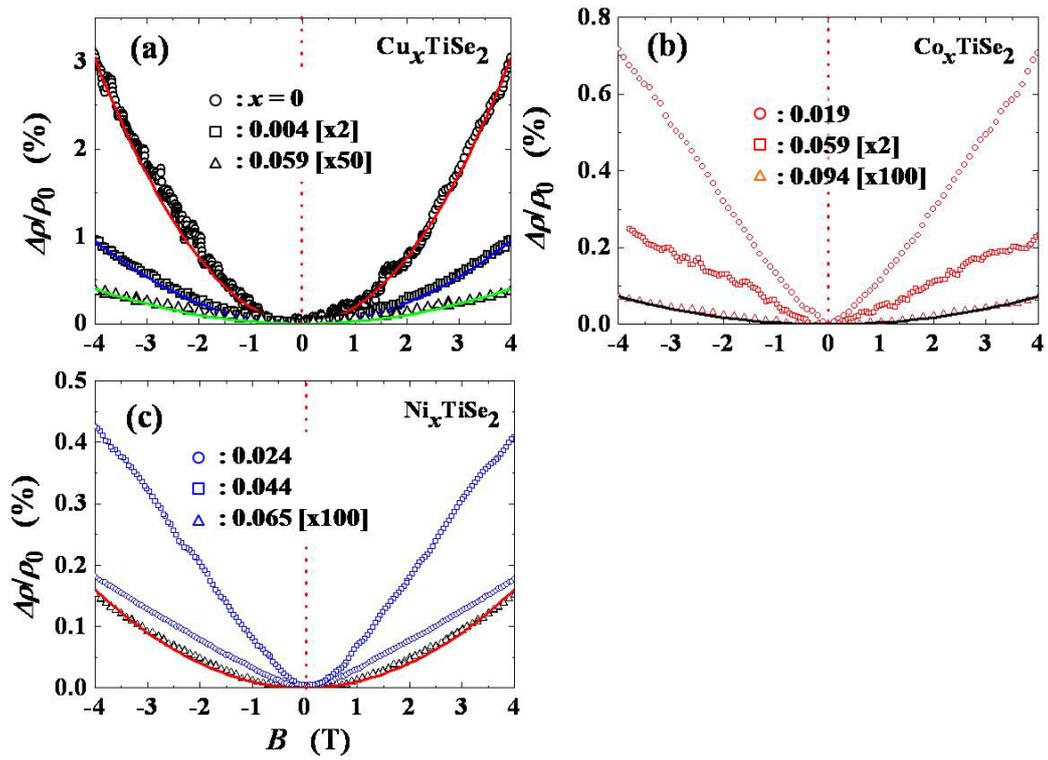

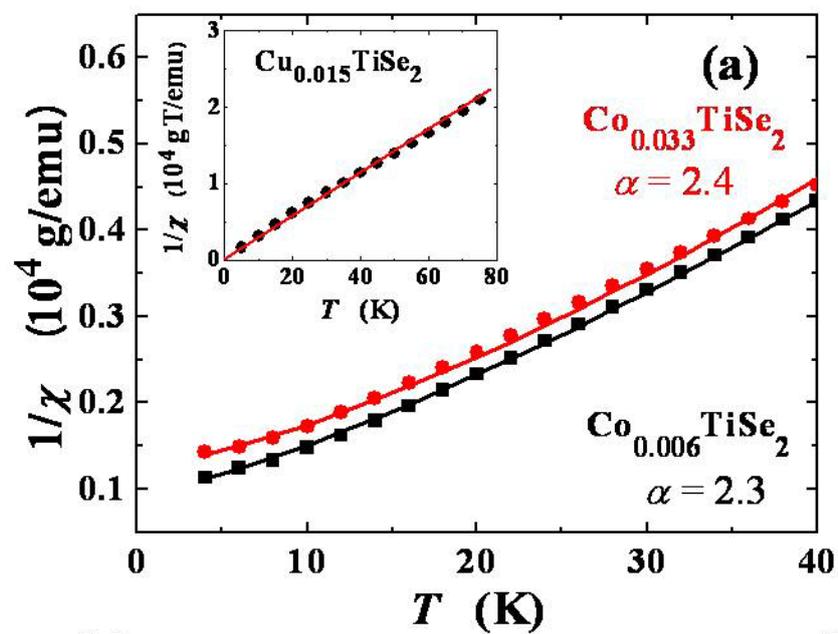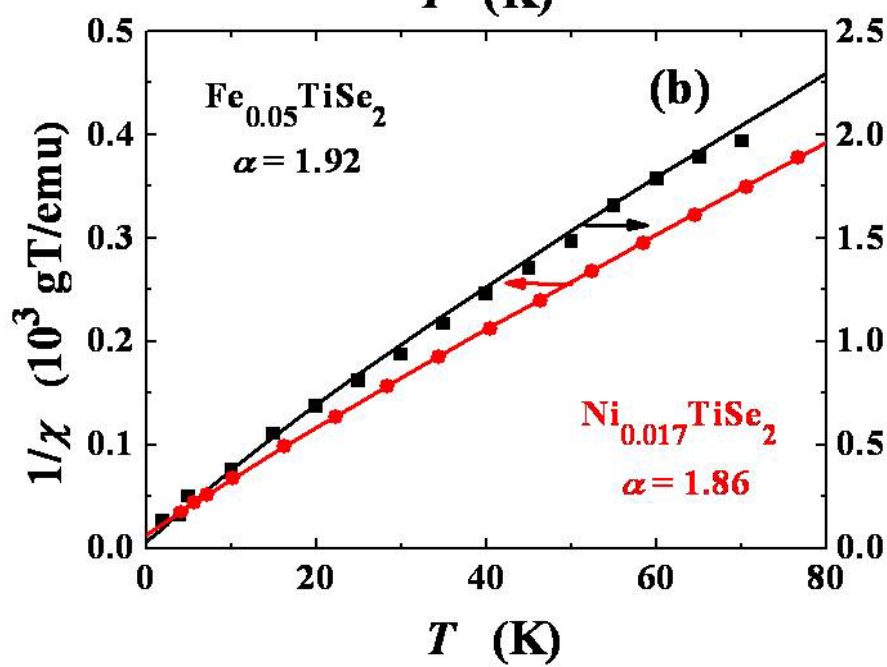

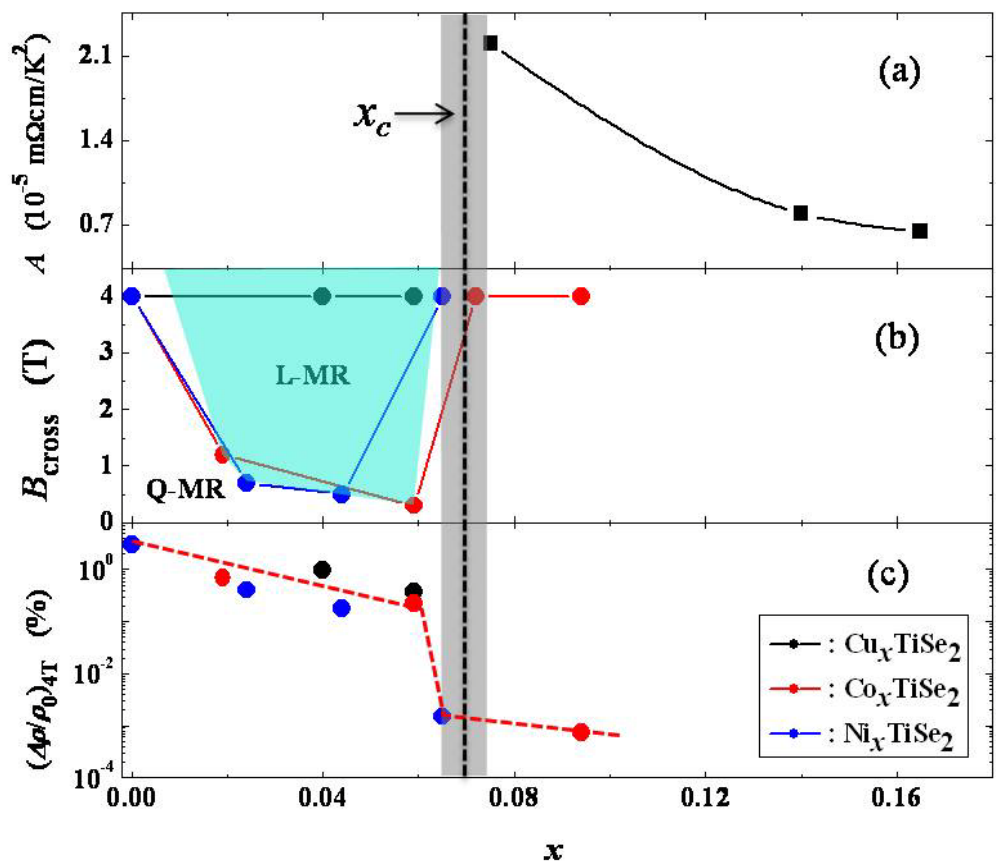

**Supplementary information for "Interplay between the Kondo effect and randomness: Griffiths phase in $M_x$TiSe$_2$ ($M$ = Co, Ni, and Fe) single crystals"**

M. Sasaki[1], A. Ohnishi[1], T. Kikuchi[1], M. Kitaura[1], Ki-Seok Kim[2], and H.-J. Kim[3*]

[1]Department of Physics, Faculty of Science, Yamagata University, Kojirakawa, Yamagata 990-8560 Japan
[2]Asia Pacific Centre for Theoretical Physics, POSTECH, Pohang, Gyeongbuk 790-784, Republic of Korea
[3]Department of Physics, College of Natural Science, Daegu University, Gyeongbuk 712-714, Republic of Korea

### Analysis method for temperature dependence of resistivity

Figure S1 shows the temperature dependence of resistivities for $M_x$TiSe$_2$ ($M$ = Cu, Co, Ni, and Fe) single crystals. In this figure, it is clearly seen that effects of magnetic ion intercalation are different from those of non-magnetic ion (Cu). In case of Cu, the peak originating in charge density wave (CDW) is drastically suppressed on the whole, moving to the lower temperature as Cu content increases. In contrast, a substantial increase of the resistivity at the low temperature is observed as indicated in the circles of Fig S1(a)-(c), while the CDW peak shows a similar behavior as Cu$_x$TiSe$_2$. The low-temperature upturn was observed to be largest in Fe$_x$TiSe$_2$. Eventually, this anomalous low-temperature upturn and the CDW peak disappears at $x > x_c$. From this observation, it is inferred that three different terms contribute to the temperature dependence of resistivity for $M_x$TiSe$_2$ ($M$ = Co, Ni, and Fe) single crystals at $x < x_c$. The first term $\rho_{bg}$ is the metallic conduction resulting from the mobile charge carriers. The second is the CDW peak and the third is the anomalous upturn at the low temperatures.

In order to isolate the anomalous upturn of resistivity, the following method was used. First, the metallic resistivity $\rho_{bg}$, whose temperature dependence is determined by electron-phonon and electron-electron scatterings at high and low temperatures, respectively is subtracted by assuming that $\rho_{bg} = \alpha \rho_r$, where $\rho_r$ is the resistivity curve above $x_c$ and $\alpha$ is the scaling parameter with nearly unity value. Since this term is the smallest among the three terms, a small variation of this term does not change our main conclusion. By using this method, we have obtained the CDW peak of Cu$_x$TiSe$_2$ as shown in the Fig. 1(a) in the main text. In fact, CDW peaks could be scaled, especially below the peak position when they are normalized by the peak values. The normalized curves are presented in Fig. S2. It can be seen that curves below the peak positions are on the same universal line. We assumed that this is a general property of the CDW peaks for $M_x$TiSe$_2$ ($M$ = Cu, Co, Ni, and Fe) single crystals.

Finally, the CDW peaks were subtracted numerically to obtain the low-temperature upturn in resistivity.

**Figure captions**

Fig. 1S (Color online) Temperature dependence of resistivity for $Fe_xTiSe_2$ (a), $Co_xTiSe_2$ (b), $Ni_xTiSe_2$ (c), and $Cu_xTiSe_2$ (d).

Fig. 2S (Color online) The scaling of the CDW parts for the $Cu_xTiSe_2$ single crystal. The temperature and the CDW contribution in resistivity are normalized with the peak values.

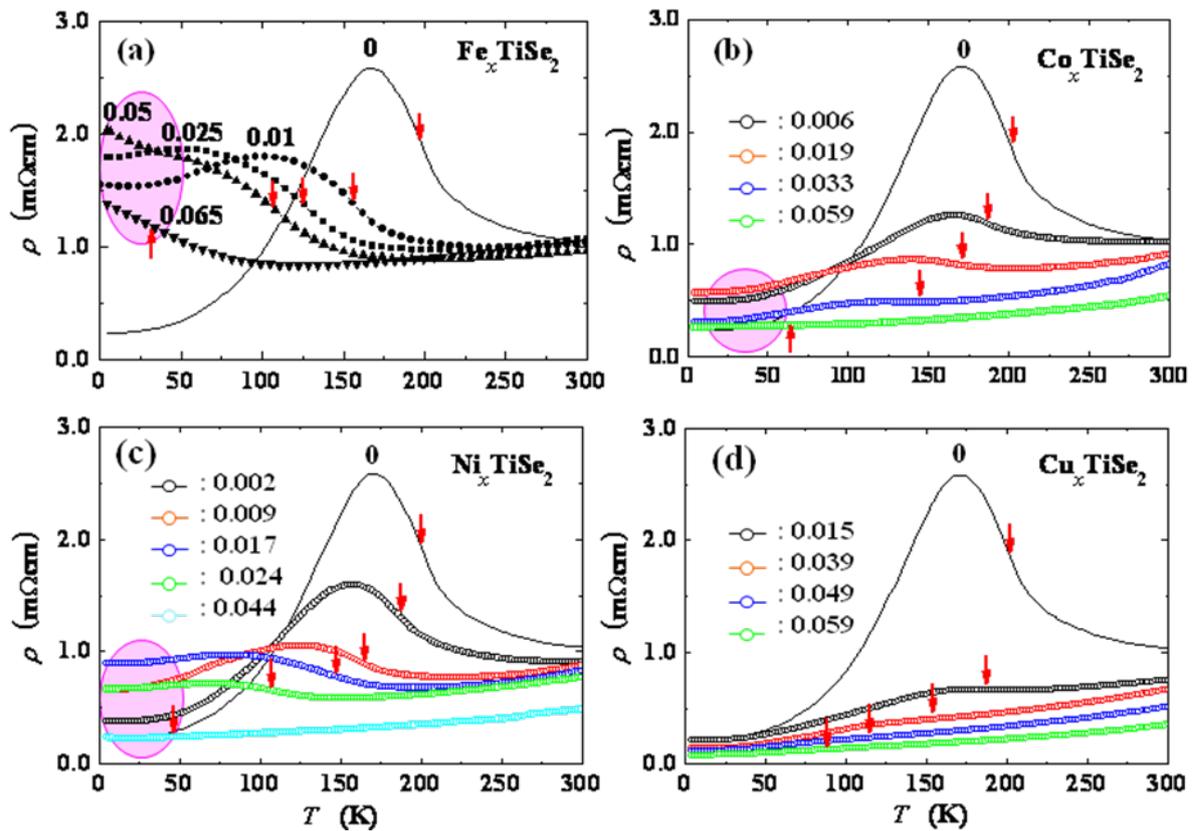

Fig. 1S

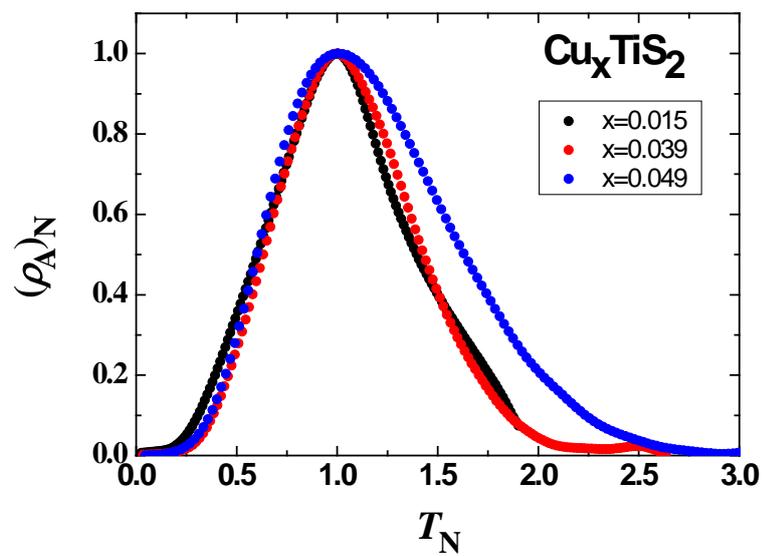

Fig. 2S